# Solution casting method for PVC nano-composites with functionalized single-walled carbon nanotubes[1]

T.V. Vlasova, V.I. Kryshtob, V.F. Mironov, L.A. Apresyan, S.N.Bokova , E.D.Obraztsova, D.V. Vlasov

Annotation

A carbon nanotubes (CNT) filled polymer nanocomposites with their unique physical, mechanical and electro-optical properties has repeatedly discussed in the literature. The main part of this problem is to obtain high quality dispersion of CNTs.
If for water-soluble polymers this problem can be considered as solved, it is not so for hydrophobic polymers, and PVC is an important example of this kind. This paper describes the methods of dispersion of SWNTs bundles in PVC, both by the use of specially synthesized surfactants, and by means of chemical surface modification of CNTs. The main criterion for an obtaining of a high degree of dispersion in our experiments is considered the sedimentation stability of non-aqueous organic colloidal solutions, which is essentially necessary for the polymer composites films preparation by solution casting method.

Как показывают теория и практика, влияние УНТ на свойства полимерных нанокомпозитов существенно зависит от характера их распределения, т.е. степени их диспергирования в объеме полимерной матрицы [1-3]. Поэтому практически полное отсутствие данных по получению и исследованию свойств полимерных нанокомпозитов на основе пластифицированного поливилхлорида (ПВХ) в качестве полимерной матрицы и различных типов углеродных нанотрубок (УНТ) в качестве гетерофазного наполнителя свидетельствует об имеющихся серьезных трудностях, связанных с разработкой эффективных способов диспергирования УНТ в случае ПВХ-матрицы. На решение этой задачи и направлено данное исследование.

Как нам представляется, наиболее удобным и эффективным в этом отношении является способ получения УНТ- нанокомпозитов методом полива из растворов. Но так как сами агрегаты УНТ в исходном состоянии практически нерастворимы ни в одном из известных растворителей, то при использовании указанного способа требуется обязательная функционализация внешней поверхности УНТ с целью получения седиментационно устойчивых их коллоидных растворов в растворителе, который одновременно является и растворителем для полимерной матрицы, т.е. способным образовывать гомогенные истинные растворы полимера.

---

[1] **Получение ПВХ-нанокомпозитов с использованием функционализированных однослойных углеродных нанотрубок методом полива из растворов**
Власова Т.В. Крыштоб В.И. Миронов В.Ф. Апресян Л.А. Бокова С.Н. Образцова Е.Д. Власов Д.В.

Для достижения этой цели, как правило, применяются методы либо физико-химической (нековалентной) модификации УНТ за счет использования специально подобранных поверхностно-активных веществ (ПАВ), либо химической (ковалентной) модификации за счет «прививки» определенных химических групп к внешней поверхности УНТ [1- 4].

В нашем случае в качестве полимерной матрицы использовался пластифицированный ПВХ, а в качестве углеродного наполнителя однослойные углеродные нанотрубки (ОУНТ). А так как получение ПВХ-ОУНТ нанокомпозитов предполагалось осуществлять методом полива из растворов, то во главу угла ставилось условие получении и использование в дальнейшем дисперсии ОУНТ в растворе ПВХ только в виде стабильных, седиментационно устойчивых нерасслаивающихся коллоидных растворов.

Для получения последних с использованием метода физико-химической (нековалентной) функционализации исходных неочищенных ОУНТ было решено:

а) использовать новый, специально синтезированный для этих целей ПАВ,

б) после проведения операции диспергирования неочищенных ОУНТ с использованием новым ПАВ в подобранном неводном органическом растворителе дополнительно провести стадию центрифугирования с целью отделения осадка и выделения для дальнейшего использования стабильного (нерасслаивающегося) седиментационно устойчивого коллоидного раствора.

С этой целью был разработан и синтезирован ПАВ-1М ( 2-гидрокси-4-додецил-6-хлорбензо[е]-1,2-оксафосфорин-2-оксид). В качестве растворителя был выбран тетрагидрофуран (ТГФ). Затем смесь расчетных количеств неочищенных ОУНТ (0,1%), ПАВ-1М (1%) и ТГФ была подвергнута ультразвуковой диспергации на установке «Сапфир-1,3» (50 Вт) в течение двух часов, после чего была проведена операция очистки полученной суспензии на центрифуге ЦЛН-1 в течение одного часа ($V_{вр}$= 7тыс.об./мин).

В итоге был получен однородный сероватого цвета седиментационно устойчивый (не расслаивающийся в течение более 3-х месяцев) коллоидный раствор очищенных ОУНТ в ТГФ. Наличие в нем ОУНТ было подтверждено спектрами комбинационного рассеяния света (КР) (см. Рис.1).

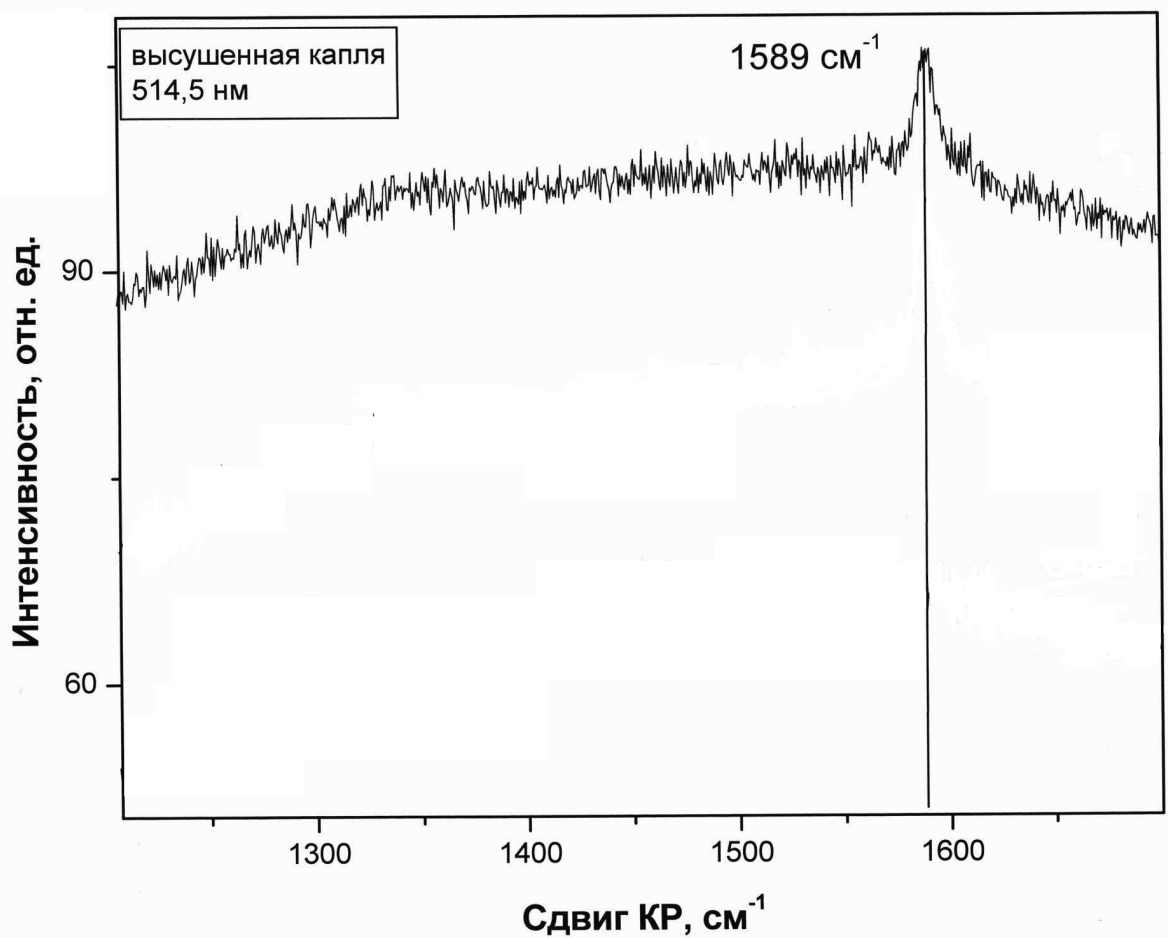

Рис.1 Спектр комбинационного рассеяния (длина волны возбуждающего излучения 514нм) высушенной капли коллоидного раствора ОУНТ с ПАВ-1М в ТГФ. Наличие ОУНТ подтверждается наблюдаемым пиком со смещением 1589 см$^{-1}$.

В дальнейшем при соединении его с 4% раствором ПВХ в ТГФ и пластифицированного одновременно пластификатором А [5] (в массовом соотношении ПВХ: модификатор А=100:50, соответственно) были получены однородные растворы, из которых потом методом полива на стеклянные подложки и последующего удаления растворителя при температуре 40$^0$С получали полимерные ПВХ-ОУНТ нанокомпозиты в виде однородных прозрачных сероватого цвета эластичных пленок.

Наряду с этим, для получения ПВХ-ОУНТ нанокомпозитов был опробован и методы химической (ковалентной) функционализации неочищенных ОУНТ. Это диктовалось, в частности, тем обстоятельством, что при использовании метода нековалентной функционализации при недостаточном соотношении ПАВ к ОУНТ образуются агрегаты, не обладающие седиментационной устойчивостью, которые быстро оседают. А при повышении концентрации ПАВ сверх необходимого для солюбилизации, возможно образование крупных полимицеллярных структур, также не обладающих седиментациоонной устойчивостью и снижающих «растворимость» комплекса ОУНТ-ПАВ [6].

Использование же методов химической (ковалентной) функционализации УНТ дает возможность перевода последних в значительно более устойчивую растворимую форму и дополнительно создает возможность использования более простых технологических методов их очистки [4].

Используя свой опыт и наработки [4], а также данные по химической «прививке» очищенных многослойных УНТ непосредственно на сам ПВХ, (методом алкилирования ПВХ нанотрубками по реакции Фриделя-Крафтса [7] ), нами была осуществлена модификация неочищенных ОУНТ методом химической функционализации макромолекулами самой ПВХ-матрицы.

Для этого брали 3-х горлую колбу с дефлегматором и магнитной мешалкой, в которую заливали 30 мл растворителя (хлороформа), добавляли при перемешивании 300 мг ПВХ марки С-70. После получения прозрачного гомогенного раствора добавляли 300мг катализатора (безводного хлорида алюминия) а затем 20 мг неочищенных ОУНТ, полученных дуговым методом. После этого смесь нагревали до $60^0$С и при перемешивании осуществляли процесс химической (ковалентной) функционализации в течение 35 часов. После завершения процесса суспензию черного цвета оставляли для отстаивания на 3 суток. Произошло четкое разделение объема всего раствора на две части: верхнюю фракцию светло-серого цвета и нижнюю – черного. Часть верхней фракции поместили в пробирку и оставили на три месяца (см. рис 2).

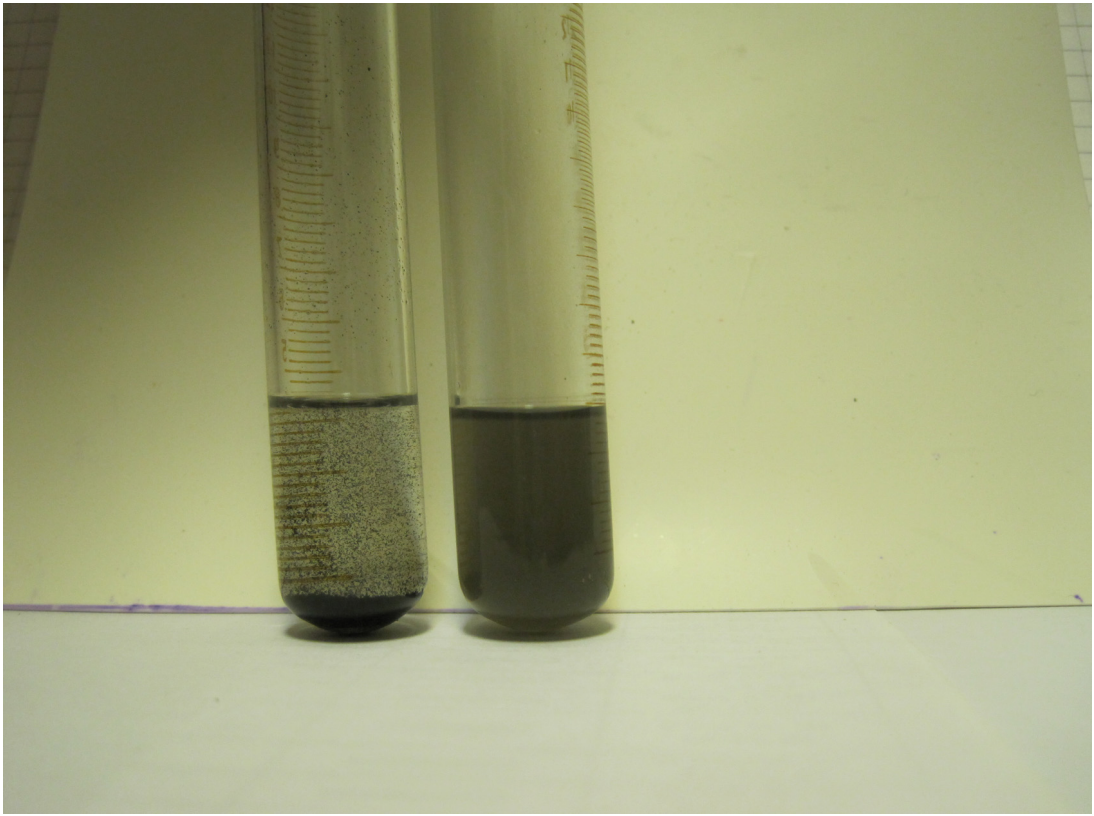

Рис2. Сравнение образцов исходной реакционной смеси для функционализации ОУНТ (слева, спустя 10мин) и коллоидный раствор функционализированных ОУНТ молекулами ПВХ (справа , спустя 3 месяца.)

При этом по истечении этого времени никаких видимых изменений (в виде появления осадка) в пробирке не наблюдалось (см. правую пробирку рис.2), из чего можно сделать вывод, что получен седиментационно устойчивый коллоидный раствор функционализированных ОУНТ, что подтверждается спектры КР (рис.3)

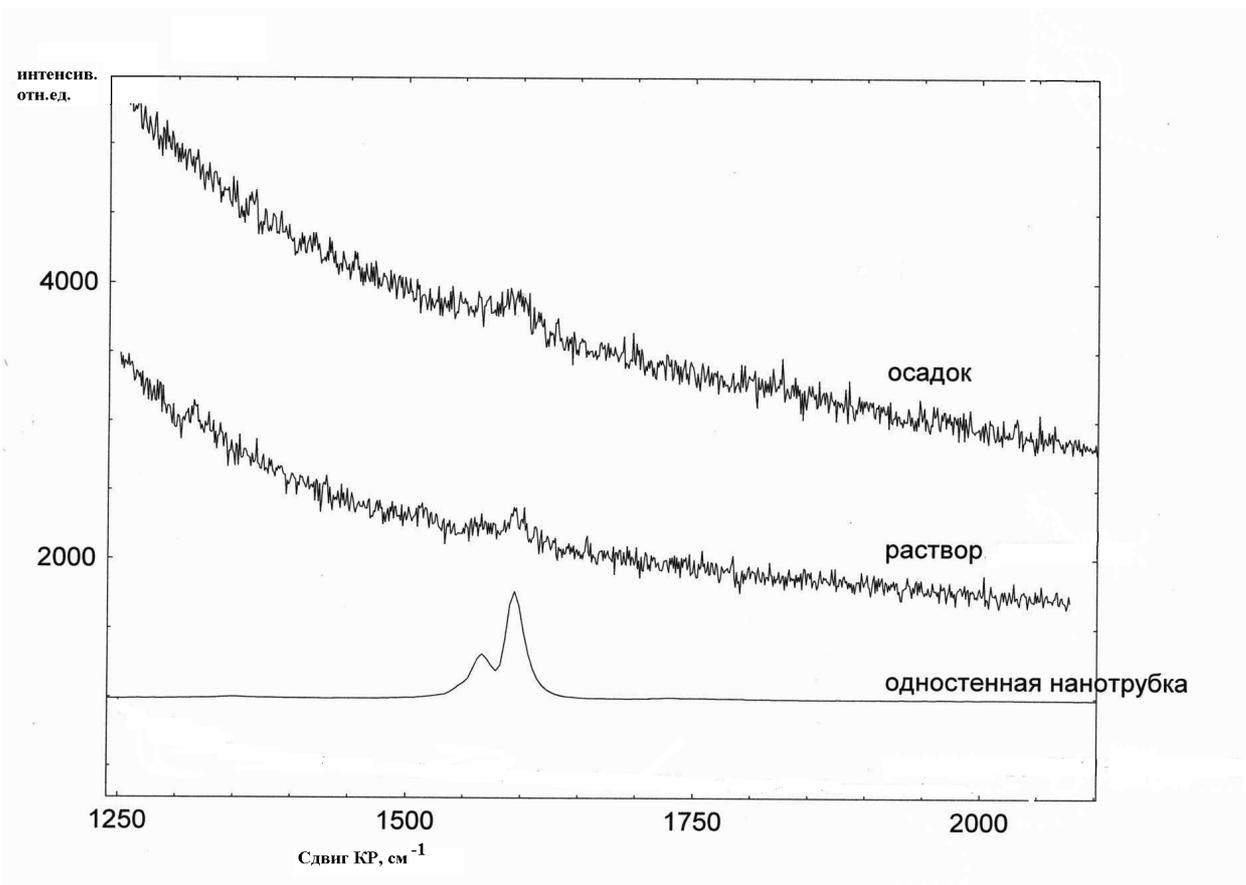

Рис.3 Спектры комбинационно рассеяния Химически модифицированных ОУНТ. Одностенная нанотрубка- эталон спектра, раствор- коллоидный раствор химически модифицированных ОУНТ, полученный после отстаивания реакционной смеси, осадок- осадок, соответственно.

Как видно при анализе спектров сигнал КР наблюдается на фоне люминесценции, которая возникает в реакционной смеси, хотя в исходном состоянии люминесценции не наблюдалось.
 Таким образом, удалось избежать стадии очистки ОУНТ методом центрифугирования.
    В дальнейшем этот коллоидный раствор был подвергнут промывке (очистке) водой от катализатора (хлорида алюминия) до отсутствия реакции на хлорид-ион с нитратом серебра в промывочных водах.
    Полученный продукт реакции высушивали и использовали для получения ПВХ-ОУНТ нанокомпозитов путем добавления и полного растворения его в 4% растворе ПВХ-матрицы в ТГФ, содержащем пластификатор А в массовом соотношении 100:25, соответственно. Образцы нанокомпозитов получали путем полива на стеклянную подложку с последующим удалением ТГФ при температуре $40^{0}C$.

Таким образом, впервые методом полива из раствора при использовании как нековалентной, так и ковалентной функционализации УНТ удалось получить ОУНТ-ПВХ нанокомпозиты в виде прозрачных серого цвета однородных эластичных пленок.